\documentclass[epj,nopacs]{svjour}
\usepackage{amsmath,amssymb,amsfonts,graphicx,cite,multirow,color,algorithmic,algorithm}
\usepackage[utf8]{inputenc}
\usepackage{amsmath}
\usepackage{caption}
\usepackage{subcaption}
\usepackage{hyperref}
\usepackage{slashed}
\usepackage[dvipsnames]{xcolor}
\graphicspath{{graphics/}}
\makeatletter
\ifx\input@path\@undefined
\def\input@path{{graphics/}}
\else
\g@addto@macro\input@path{{graphics/}}
\fi
\makeatother
\newcommand{\mg}{MG5aMC}
\newcommand{\mga}{\textsc{MadGraph5\_aMC@NLO}}

\input{def.sty}

\preprint{LU-TP 22-19\\MCNET-22-05}

\title{Automating scattering amplitudes with chirality flow}

\author{Andrew Lifson\inst{1}, Malin Sj\"odahl \inst{1} and Zenny Wettersten\inst{1}}

\institute{
  Department of Astronomy and Theoretical Physics, Lund
  University, Box 43, 221 00 Lund, Sweden}

\date{\today}

\abstract{
  Recently we introduced the chirality-flow formalism, a method which builds
  on the spinor-helicity formalism and is inspired by the color-flow idea in
  QCD. With this formalism, Feynman rules and diagrams are simplified to the
  extent that it is often possible to immediately, by hand, write down a 
  helicity amplitude given a Feynman diagram.
  In this paper we show that the method can also speed up numerical evaluation
  of scattering amplitudes by considering $e^+ e^-$ going to $n$ photons in
  a \textsc{MadGraph}-based tree-level implementation.  We find that the computation time is 
  reduced by roughly a factor ten for six photons,  and that it scales better 
  with the number of external particles than the default \mga{} implementation. This performance gain is in part attributed 
  to the more compact Lorentz structures involved,  and in part due to a transparent choice of
  gauge reference vectors which reduces the number of Feynman diagrams considered.
}

\begin{document}

\maketitle

\section{Introduction}
\label{sec:introduction}

Event generators \cite{Gleisberg:2008fv,Alwall:2014hca,Sjostrand:2014zea,Bellm:2015jjp,Sherpa:2019gpd}
for scattering amplitudes are indispensable tools for calculating
cross sections and understanding event topologies at collider experiments.

At the core of amplitude calculations is the evaluation of
the hard scattering matrix element, typically calculated using Feynman
diagram techniques as helicity amplitudes \cite{Murayama:1992gi,Gleisberg:2008fv,Alwall:2014hca},
i.e., amplitudes with assigned helicities.

While such calculations may well be performed using the full
four-dimensional Dirac spinors, simplifications can be achieved
using the spinor-helicity \cite{DeCausmaecker:1981jtq,
Berends:1981rb,Berends:1981uq,DeCausmaecker:1981wzb,
Berends:1983ez,Kleiss:1984dp,Berends:1984gf,Gunion:1985bp,
Gunion:1985vca,Kleiss:1985yh,Hagiwara:1985yu,Kleiss:1986ct,
Kleiss:1986qc,Xu:1986xb,Gastmans:1987qz,Schwinn:2005pi}
and Weyl-van der Waerden \cite{Farrar:1983wk,Berends:1987cv,
Berends:1987me,Berends:1988yn,Berends:1988zn,Berends:1989hf,
Dittmaier:1993bj,Dittmaier:1998nn,Weinzierl:2005dd,Gleisberg:2008fv}
formalisms,  in which spinors are decomposed into their 
{\blue{left}}- and {\red{right}}-chiral
parts which transform separately as different
$SL(2,\mathbb{C})$ copies (see e.g.\  \cite{Mangano:1990by,
Dixon:1996wi,Dreiner:2008tw, Elvang:2013cua,Dixon:2013uaa} 
for pedagogical introductions).   The Dirac spinors are thus split into 
\begin{align}
u(p) &\sim v(p) \sim \begin{pmatrix}
\sqrSp{p_1} \\
\ranSp{p_2}
\end{pmatrix}\;,
&
\ubar(p) &\sim \vbar(p) \sim \begin{pmatrix}
\sqlSp{p_1} \,\,, & \; \lanSp{p_2}
\end{pmatrix}\nonumber
~,
\end{align}
for some $p_1$ and $p_2$, one of which reduces to $p$ in the
ultrarelativistic/massless case,  while the other vanishes.
\footnote{
While this paper deals with massless particles, massive fermions
can easily be treated. For example,  an outgoing spinor of positive helicity
and mass $m$ is $\ubar^+(p) = \begin{pmatrix}
[p_f| \,\,, & \frac{m}{\langle p_b p_f \rangle}\langle p_b |
\end{pmatrix}$, where $p_f$ and $p_b$ 
are the forward and backward components of the momentum $p$,
i.e., $p_{f/b}=\frac{p^0\pm|\vec{p}|}{2}(1,\pm \hat{p})$
 \cite{Kleiss:1985yh,Dittmaier:1998nn,Weinzierl:2005dd,Alnefjord:2020xqr}.
} 

Here the \Lcolour{square} brackets are Weyl spinors transforming
under $\Lcolour{SL(2,\mathbb{C})_{\Lcolour L}}$ and the \Rcolour{angled}
brackets are Weyl spinors transforming under $\Rcolour{SL(2,\mathbb{C})_{R}}$.

Since the only invariant tensor for $SL(2,\mathbb{C})$
is the fully antisymmetric
Levi-Civita tensor,
$\epsilon^{12} = -\epsilon^{21} = \epsilon_{21} = -\epsilon_{12} = 1$,
invariant spinor inner products are formed by
contractions with this tensor
\begin{eqnarray}
      \underbrace{\epsilon^{\al \be } \ranSp{i}_{\be}}_{\equiv \lanSp{i}^{\al}}\ranSp{j}_{\al}
      &=&\lanSp{i}^{\al}\ranSp{j}_{\al}
      =\lan i j \ran=-\lan j i \ran,
      \nonumber\\
      \underbrace{\epsilon_{\da \db} \sqrSp{i}^{\db}}_{\equiv \sqlSp{i}_{\da}} \sqrSp{j}^{\da}
      &=& \sqlSp{i}_{\da} \sqrSp{j}^{\da}
      =\sql ij\sqr =-\sql j i\sqr\;,\nonumber
\end{eqnarray}
where we denote $\lanSp{p_i} = \lanSp{i}$ etc., for brevity,
and where (up to a phase)
$\lan ij \ran \sim \sql ij \sqr \sim \sqrt{2p_i\cdot p_j}\;$.

Since these are the only invariant structures at hand,
it can be anticipated that all scattering amplitudes should be
expressible in terms of these spinor inner products,
and that depicting the contraction with a connecting line,
one can obtain a ``flow'' picture for the Lorentz structure.

This flow picture, the chirality-flow formalism is
introduced in the next section, along with an illuminating example
of how to write down amplitudes. In the subsequent section,
\secref{sec:implementation}, we describe our chirality-flow
implementation based on the \mga{} framework \cite{Alwall:2014hca}. 
After that, the obtained speed gain
is digested in \secref{sec:results}. Finally,
concluding remarks and an outlook are given in
\secref{sec:conclusion}.

\section{Chirality flow}
\label{sec:chirality flow}

In the chirality-flow formalism \cite{Lifson:2020oll,Lifson:2020pai, Alnefjord:2020xqr,Alnefjord:2021yks} \,
we take the simplifications of the spinor-helicity formalism one step further.
By proving that we can recast Feynman rules to be represented in
terms of flows between external spinors, we manage to simplify Feynman rules
and diagrams to the extent that helicity amplitudes can often be
immediately written down given a Feynman diagram.

Introducing graphical flow representations \cite{Lifson:2020pai} for the external
spinors, 
\begin{alignat}{2}
 \lanSp{i} &= \raisebox{-0.3\height}{\includegraphics[scale=0.4]{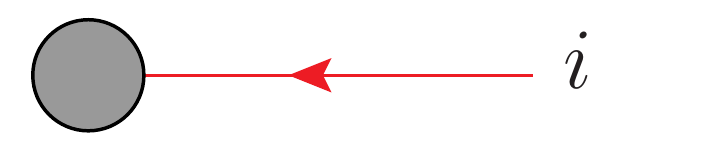}},
 \qquad  
   \sqlSp{i} &&= \raisebox{-0.3\height}{\includegraphics[scale=0.4]{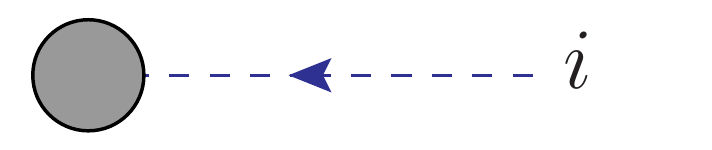}},
 \nonumber \\
 \ranSp{j} &= \raisebox{-0.3\height}{\includegraphics[scale=0.4]{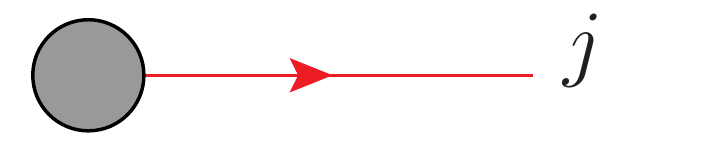}}, 
   \qquad 
   \sqrSp{j} &&= \raisebox{-0.3\height}{\includegraphics[scale=0.4]{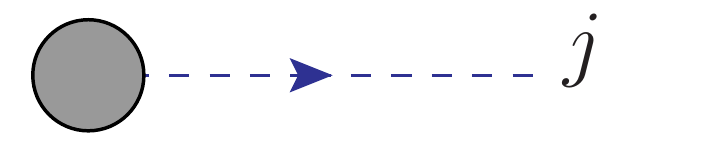}} ,
   \nonumber
\end{alignat}
we can --- in analogy with the color-flow representation of gluons in QCD --- 
obtain a double line representation for external spin-1 particles.
Letting $\epsilon_{\blue{L}}(p_i,r)$ denote a left-chiral (negative helicity incoming or
positive helicity outgoing) photon of momentum $p_i$ and with gauge reference
vector $r$, and similarly $\epsilon_{\red{R}}(p_i,r)$ denote a right-chiral
(positive helicity incoming or negative helicity outgoing)
photon, we have
\begin{eqnarray}
  \epsilon_{\blue{L}}(p_i,r) &\rightarrow& \frac{1}{\lan r i \ran}\raisebox{-0.2\height}{\includegraphics[scale=0.35]{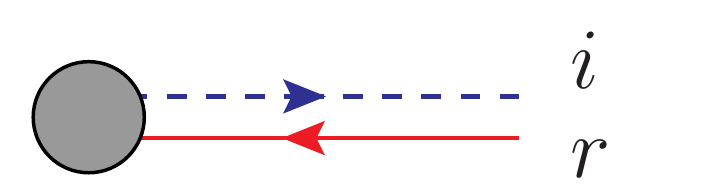}}\text{or}
  \quad \frac{1}{\lan r i \ran} \raisebox{-0.2\height}{\includegraphics[scale=0.35]{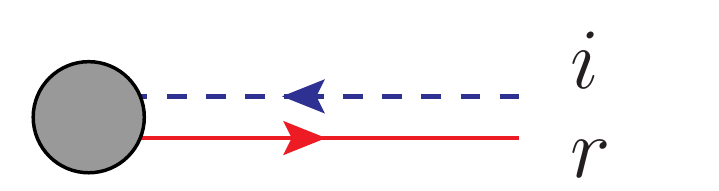}}\hspace*{-2 mm},\nonumber\\
  \epsilon_{\red{R}}(p_i,r) &\rightarrow& \frac{1}{\sql i r \sqr}\raisebox{-0.2\height}{\includegraphics[scale=0.35]{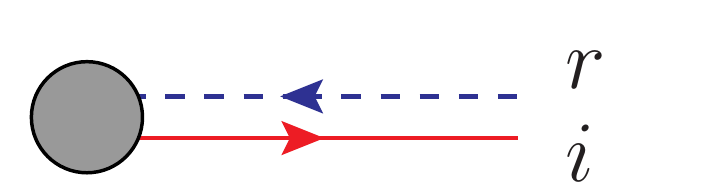}}\text{or}
  \quad \frac{1}{\sql i r \sqr} \raisebox{-0.2\height}{\includegraphics[scale=0.35]{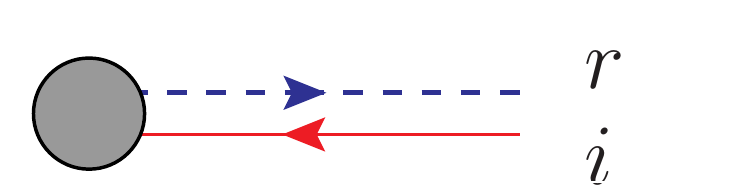}}\hspace*{-2 mm}.\nonumber
\end{eqnarray}
Note that the unphysical reference momentum $r$ is carried by
the right-chiral line for a left-chiral photon, and vice versa. 

In \cite{Lifson:2020pai}, we proved that we can always use the Fierz identity
\begin{equation}
  \lanSp{i} \taubar^\mu \sqrSp{j} \sqlSp{k} \tau_\mu \ranSp{l} 
    = \lan il \ran \sql kj \sqr
    \nonumber
\end{equation}
on Dirac matrices decomposed into the Pauli matrices\footnote{
This normalization of the Pauli matrices is chosen to avoid carrying and canceling unnecessary factors of $\sqrt{2}$.
}, $\tau^\mu=\sigma^\mu/\sqrt{2}$,
combined with charge conjugation
(see e.g.\ \cite{Dixon:1996wi,Elvang:2013cua})
\begin{equation}
  \lanSp{i} \taubar^\mu \sqrSp{j} =\sqlSp{j} \tau^\mu \ranSp{i}
  \nonumber
\end{equation}
to replace a photon (spin-1) propagator by a solid and a
dotted line with arrows opposing
\begin{eqnarray}
  &&-i\frac{g_{\mu\nu}}{p^2}
  \;\;=\; 
  \raisebox{-0.25\height}{\includegraphics[scale=0.4]{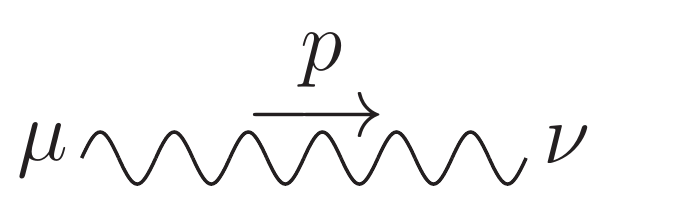}}
  \rightarrow \nonumber\\
  &&-\frac{i}{p^2}\raisebox{-0.25\height}{\includegraphics[scale=0.4]{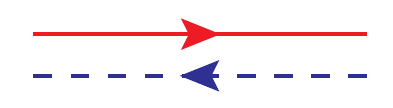}}
  \quad \mbox{or}\quad
  - \frac{i}{p^2} \ \raisebox{-0.25\height}{\includegraphics[scale=0.4]{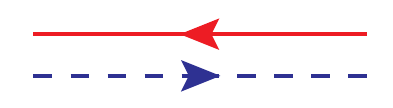}}\;. \nonumber
  \label{eq:vbpropagator}
\end{eqnarray}

The arrow direction, for internal as well as external photons,
has to be chosen such that the arrows in the diagram align with
each other (rather than oppose each other).

This also enables us to recast the fermion-photon (spin-1)
vertex into a simple flow form
\begin{eqnarray}
  \label{eq:fermion_photon_vertex0}
  ie\sigma^{\mu}=\raisebox{-0.35\height}{\includegraphics[scale=0.4]{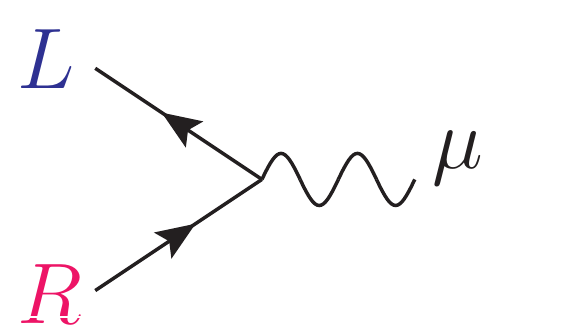}}
  & \rightarrow&
  ie\sqrt{2}\raisebox{-0.4\height}{\includegraphics[scale=0.35]{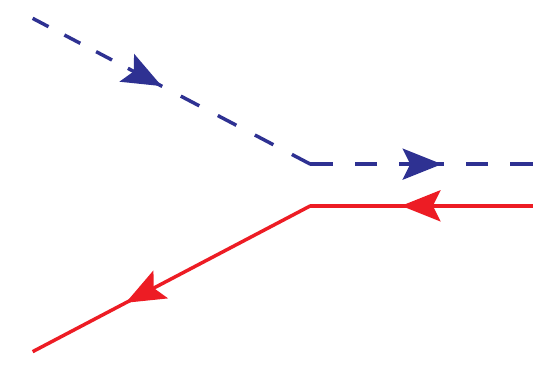}}\nonumber\;, \\
  ie\sigmabar^{\mu}=\raisebox{-0.35\height}{\includegraphics[scale=0.4]{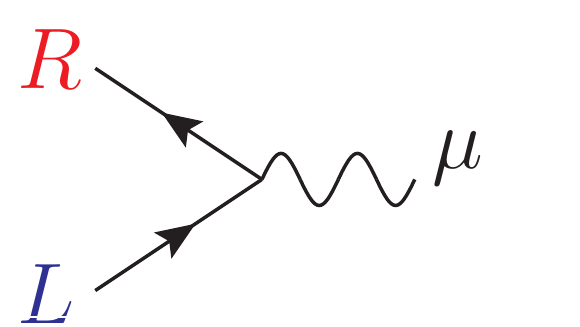}} 
  & \rightarrow&
  ie\sqrt{2}\raisebox{-0.4\height}{\includegraphics[scale=0.35]{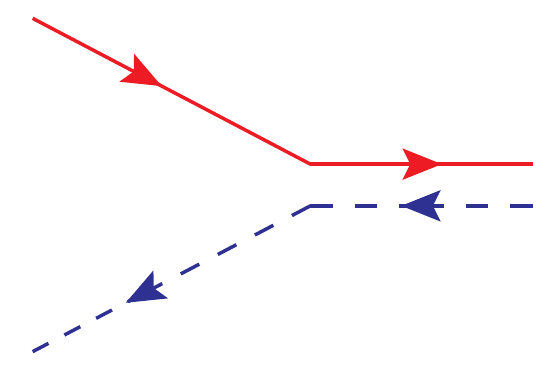}}\;.
  \nonumber
\end{eqnarray}

The fermion propagator requires some more consideration, but
the parts contracted with $\sigma$ and $\bar{\sigma}$ can be
represented graphically by
\begin{eqnarray}
  \label{eq:fermion_propagator}
  \frac{i}{p^2}p_\mu \sigma^\mu=\frac{i}{p^2}\slashed{p}
  &\rightarrow&
  \frac{i}{p^2}
  \raisebox{-0.10\height}{\includegraphics[scale=0.4]{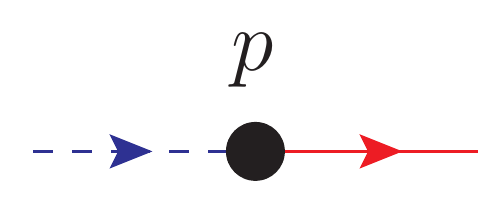}}
\quad  \text{and} 
  \nonumber \\
  \frac{i}{p^2}p_\mu\bar{\sigma}^\mu=\frac{i}{p^2}\bar{\slashed{p}}
  &\rightarrow&
  \frac{i}{p^2}
  \raisebox{-0.10\height}{\includegraphics[scale=0.4]{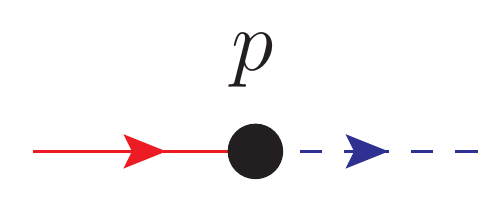}},\nonumber
  \label{eq:fermion_propagator2}
\end{eqnarray}
respectively.

Decomposed into massless momenta $p_i$, with $p=\sum_i p_i$, $p_i^2=0$,
we have for the first term
\begin{eqnarray}\label{eq:fermion_prop_decomp}
  \quad \raisebox{-0.15\height}{\includegraphics[scale=0.4]{./Jaxodraw/MS/FermionMomDotaCol}}
  \;
  &=
  \;\sum_i\; \raisebox{-0.20\height}{\includegraphics[scale=0.4]{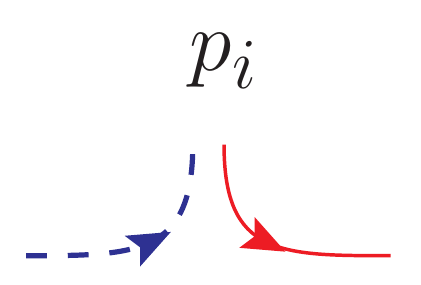}}\;, \nonumber
\end{eqnarray}
and similar for the second term.

Applying these rules, it is possible to
directly write down scattering amplitudes, either in
terms of slashed momenta or in terms of
Lorentz-invariant spinor inner products.

Calculations with Feynman diagrams can then be simplified in an
unprecedented manner, making them trivial. To illustrate this,
we consider a Feynman diagram relevant for $e^+_Re^-_L\rightarrow n$ photons,
and overlay the chirality-flow representation
\begin{equation}
 {}\raisebox{-0.50\height}{\includegraphics[scale=0.4]{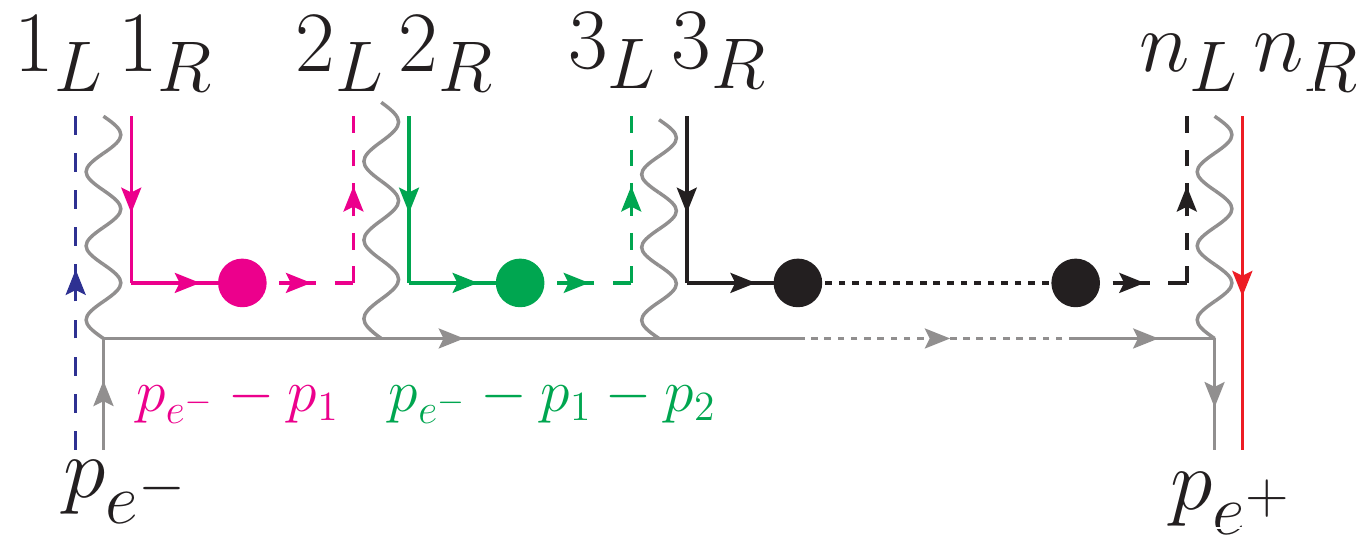}}
 \nonumber.
\end{equation}
Here, for a left-chiral photon $1_L=p_1$ and $1_R=r_1$, and vice versa for a
right-chiral photon, and we have used the freedom to assign chirality-flow
arrows in any consistent direction.

Writing down the amplitude (as an example for photon 1 right-chiral, and
photons 2 and 3 left-chiral), either in terms of $\sigma/\sigmabar$
matrices,
\begin{eqnarray}
&&   \underbrace{(\sqrt{2} i e)^{n}}_{\text{from vertices}}
  \underbrace{\frac{(i)^{n-1}}{(p_{e^-}-p_1)^2\;\hdots}}_{\text{from fermion propagators}} \
  \underbrace{\frac{1}{ [1\, 1_L ]  \langle 2_R\, 2 \rangle\langle 3_R\, 3 \rangle\hdots} }_{\text{from polarization vectors}}
  \nonumber \\
&& \times {\textcolor{blue}{[p_{e^-}\,1_L  ]}}
    \times {\textcolor{magenta}{
        \langle  1_R|\bar{\slashed{p}}_{e^- - 1}  |2_L  ] 
    }}\times 
    {\textcolor{ForestGreen}{
        \langle 2_R| \bar{\slashed{p}}_{e^- - 1 - 2}|3_L]
        }}
    \nonumber  \\
    && \times \hdots \times 
    {\textcolor{red}{  \langle  n_R\,p_{e^+}  \rangle} }\;,
    \label{eq:fermionPropSigma}
\end{eqnarray}
or with the spinor structure directly expressed in terms of spinor inner products,
\begin{eqnarray}
  &&
    {\textcolor{blue}{[p_{e^-}\,1_L  ]}}
    \times {\textcolor{magenta}{\Big(
        \langle  1_R\,p_{e^-}  \rangle [p_{e^-}\,2_L  ]
       -\langle  1_R\,1  \rangle [1\,2_L  ] \Big)
    }}\nonumber \\
    && \times 
    {\textcolor{ForestGreen}{\Big(
        \langle 2_R\, p_{e^-}\rangle [p_{e^-}\,3_L]
        - \langle 2_R\, 1\rangle [1\,3_L]
        - \langle 2_R \,2\rangle [2\,3_L]
        \Big)}}
    \nonumber \\
    && \times
    \hdots \times 
    {\textcolor{red}{  \langle  n_R\,p_{e^+}  \rangle} }\;,
    \label{eq:res}
\end{eqnarray}
we see that this diagram vanishes if the reference momentum
$r_1=1_L$ is chosen to be $p_{e^-}$, since ${\textcolor{blue}{[p_{e^-}\,p_{e^-} ]}=0}$,
i.e. for a right-chiral photon, this diagram
can be chosen to disappear. By picking the gauge vector
to be $p_{e_-}$ for all right-chiral photons, we can make
all diagrams with a right-chiral photon attached directly to the electron disappear.
Similarly, by letting the reference momentum be $p_{e^+}$
for all left-chiral photons, diagrams with a left-chiral
photon attached next to $p_{e^+}$ vanish.

If a given assignment of photon chiralities has $n_L$ 
left-chiral photons and $n_R$ right-chiral photons,
then we have $n_L$ non-vanishing ways of placing a photon next to the electron,
$n_R$ ways to place a photon next to the positron,
and $(n-2)!$ possible ways to order the remaining photons.
This leaves us with $n_L n_R (n-2)!$, rather than $n!$ diagrams
to consider for this chirality assignment, 
a simplification which turns out to reduce
the computation time significantly.

By consistently using this gauge choice, diagram generation 
can be constructed to recognize any vertices coupling 
a left-chiral (right-chiral) photon with the right-chiral (left-chiral) 
fermion as not contributing to the amplitude,
and the diagrams can be removed already before compile time.
We refer to this process as \textit{gauge based diagram removal}.

We note that the same simplification could have been
achieved within the spinor-helicity or 
Weyl-van der Waerden formalisms, but with chirality flow it is
completely transparent.

\section{MadGraph implementation}
\label{sec:implementation}

To test the viability of a numerical implementation of 
chirality flow,  we create a UFO \cite{Degrande:2011ua} model with chiral particles and vertices, 
feed this into the 
software framework \mga (\mg)\cite{Alwall:2014hca} in standalone mode,  
and repurpose the amplitude evaluations to work within
the chirality-flow formalism.

To make the current helicity amplitude evaluation and our implementation
as comparable as possible, we follow the structure of \mg{} wherever possible. 
We therefore only 
1) modify \mg{}'s diagram generation in order to produce 
chirality-flow diagrams, and 2) replace the underlying 
numerical  HELAS-like routines generated by ALOHA \cite{deAquino:2011ub} 
for calculating off-shell currents and amplitudes with  
a similar library performing these calculations based on chirality flow.

Although this implementation does not lend itself 
immediately to all the possible benefits of the 
chirality-flow formalism, it does make runtime 
comparisons as fair as possible\footnote{
Since we use standalone output, 
some optimizations such as the recently implemented helicity recycling 
\cite{Mattelaer:2021xdr} are not included in the \mg{} 
speed (though this recycling will in theory equally apply to chirality flow).  
Rather,  the comparison we make singles out the gains due to simpler 
Lorentz structures and gauge based diagram removal.},   
as we are performing the same type of evaluations of 
the same type of processes using the same type 
of program. Any advantage in evaluation time will 
thus be due to simplified calculations (involving smaller Lorentz structures)
or a reduced \textit{number} of evaluations (as for gauge based
diagram removal).

The evaluation process performed in our 
implementation is identical to the \mg{} version, 
although with explicitly chiral particles and vertices. 
As \mg{} treats different helicity states of a 
particle as the same type of particle, this means that 
our implementation runs what in \mg{} is a helicity state 
as its own process. If we wish to perform a helicity summed calculation, 
as in standalone \mg{}, we need to run each chirality 
configuration as its own subprocess. This brings about 
some overhead\footnote{As each helicity configuration is 
computed as a distinct process, phase space points are generated 
independently for each process using the RAMBO algorithm 
\cite{Kleiss:1985gy}. For the simplest process $e^+ e^- \to 2\gamma$,
an analysis with Valgrind \cite{Nethercote:2007vaf,Weidendorfer:2004ccs} 
shows that RAMBO ends up taking roughly half of 
the runtime for our implementation, whereas the time used by RAMBO
is negligible for many photons.}.

Aside from generating chirality-flow diagrams, rather 
than standard Feynman diagrams, we also replace the 
libraries for the matrix element evaluations 
with a library performing the corresponding 
calculations using chirality flow. However, as \mg{} 
stores particle momenta locally at each vertex evaluation, 
the decomposition of the fermion propagator momentum is 
based on \eqref{eq:fermionPropSigma} rather than \eqref{eq:res}.
Changing this and implementing caching of spinor inner products
could likely offer additional speed gain.

To validate our implementation, 
several classes of processes were evaluated at random 
points in phase space and compared with the same 
processes evaluated at the same phase space points 
using default \mg{}. 
The processes validated include $e^+ e^- \to n \gamma$,
$2\leq n \leq 4$;
$e^+ e^- \to \mu^+ \mu^- n\gamma$,
$0\leq n \leq 2$;
$e^+ e^- \to e^+ e^- n\gamma$,
$0\leq n \leq 2$;
$e^-\gamma \to e^- n\gamma$,
$1\leq n \leq 3$;
$e^+ e^- \to 2 \mu^+ 2\mu^-$; $e^+ e^- \to e^+ e^- \mu^+ \mu^-$; 
and $e^- \mu^- \to 2e^- \mu^- e^+$, for all possible helicity
configurations. 
All amplitudes were found to be equal within 
numerical precision.

\section{Results}
\label{sec:results}

\begin{figure}[t]
    \centering
    \includegraphics[width=0.45\textwidth]{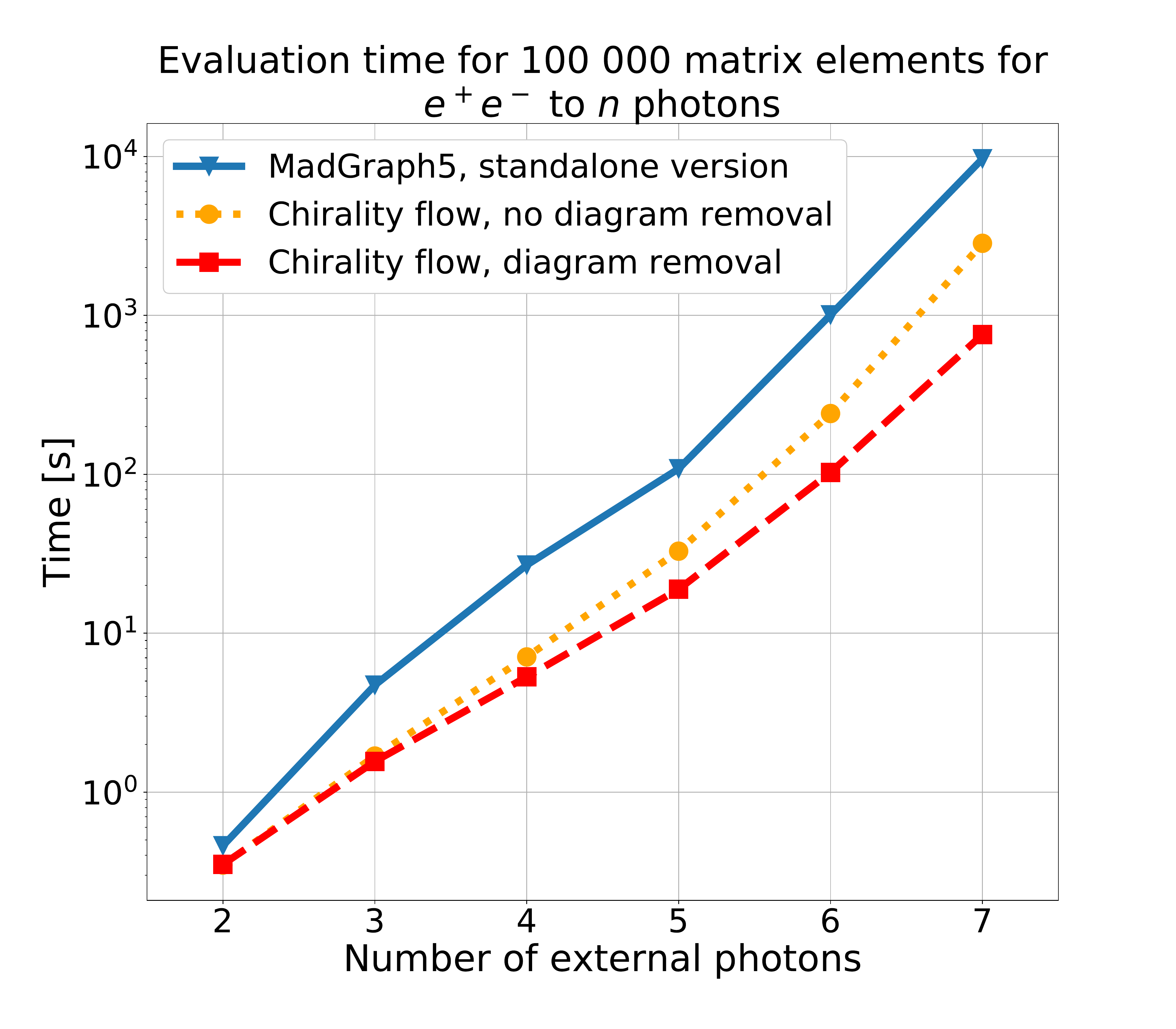}
    \caption{Measured runtimes for evaluation of 
    100 000 matrix elements for \mg{} and our implementation, 
    as a function of photon multiplicity $n$. The dotted orange 
    curve depicts the chirality-flow implementation 
    without gauge based diagram removal, whereas the dashed 
    red curve includes gauge based diagram removal. 
    The same number of helicity/chirality configurations 
    are evaluated in all cases. The chirality-flow implementations 
    evaluate matrix elements faster than \mg{} for 
    all $n$. 
    For small $n$ the gain can be explained almost 
    fully by simplified calculations, 
    but for larger $n$ the effects of 
    gauge based diagram removal become readily apparent.
    }
    \label{fig:EvaluationTime_real}
\end{figure}

Figure \ref{fig:EvaluationTime_real} depicts measured 
runtimes for evaluation 
of 100 000 matrix elements for the 
process $e^+ e^- \to n\gamma$ as a function of 
photon multiplicity $n$ (measured on an AMD Ryzen 5 1600 CPU).
Three implementations are depicted: 
\mg{} (solid blue line), our implementation without 
gauge based diagram removal (dotted orange line), 
and our implementation with gauge based diagram removal 
(dashed red line).
These matrix elements were evaluated 
for phase space points generated by RAMBO \cite{Kleiss:1985gy}, 
all using the same seed.
The comparison of the dotted and solid lines shows the
improved evaluation speed obtained by performing calculations 
using the simplified Lorentz structure, whereas the 
difference between the dotted line and the dashed 
line is due to gauge based diagram removal.

For this comparison, \mg{} has been manually set 
to consider only contributing helicity 
configurations, and subprocesses 
where all external photons have the same chirality 
have been discarded for the implementation without 
diagram removal\footnote{
\mg{} standalone has a routine for 
detecting non-contributing helicity configurations. 
For the process we consider,  this routine discards configurations where the 
fermions have the same helicity, but it does not 
discard configurations where all photons have the 
same helicity (using \textsc{MadGraph5\_aMC@NLO} 3.2.0,
with standalone output, 
released on August 22nd, 2021).
}.
With this setup, all three program versions 
will evaluate the same number of matrix elements for 
the same number of helicity/chirality configurations. 

As can be seen in Figure \ref{fig:EvaluationTime_real}, 
the chirality-flow implementations perform the evaluations faster than
\textsc{Mad-}\\\textsc{Graph5}
for all photon multiplicities $n$. Additionally, 
for small $n$ both chirality-flow versions scale 
better with $n$ than \mg{}, and chirality flow 
with gauge based diagram removal maintains this 
gentler slope with $n$ into the region of large $n$.

In the small $n$ region, $n \lesssim 4$, the difference 
between \mg{} and chirality flow is explained almost 
entirely by the simplified Lorentz structures, but as the large $n$ 
region is approached, $n \gtrsim 5$, the benefit 
of gauge based diagram removal becomes clear. 
At $n=6$, chirality flow with diagram removal 
is roughly a factor 10 faster than \mg{}.

\section{Conclusion and outlook}
\label{sec:conclusion}

In previous articles, we have developed chirality flow 
and shown its benefits for analytic 
calculations. Here, we have further demonstrated the 
viability of the chirality-flow formalism in 
a numerical \mg{} implementation.

As Figure \ref{fig:EvaluationTime_real} demonstrates, 
our chirality-flow based implementation evaluates 
matrix elements faster than standalone \mg. This speed increase 
is due to two different factors. The first is the 
simplified calculations obtained by performing 
evaluations of Lorentz structures in the 
chirality-flow formalism. 
The second, however, is an additional benefit of 
chirality flow: the effect of gauge reference vector choice becomes very transparent. 
Since the effects of a given reference momentum can 
be seen directly from the chirality-flow diagrams,
a good choice is immediately discernible. 
Combining these two effects, the process $e^+ e^- \to n \gamma$ 
ends up being evaluated roughly ten times faster for $n \geq 6$
in our implementation, with speed gain increasing for an increasing
number of photons.

In this paper, the \mg{} structure has been maintained for 
the purpose of comparison. However, the HELAS-based structure 
used by \mg{} is not naturally suited for chirality flow,
since it evaluates diagrams by calculating 
off-shell particle wavefunctions at vertices, 
before combining them to calculate an amplitude. 
This general structure allows recycling of currents 
for each diagram where the given current enters.

For chirality flow, another natural object to recycle
is the spinor inner products (cf. \eqref{eq:res}). 
Since the amplitude corresponding to a given chirality-flow 
diagram can be expressed by a small number of these scalars, 
caching them is likely to increase the evaluation speed even further 
in future implementations.

While this implementation has concerned only massless QED, the
chirality-flow formalism has been developed for the full massive
Standard Model \cite{Alnefjord:2020xqr},  and sizable speed gains could likely be 
attained for a large class of phenomenologically
relevant Standard Model processes.

\section*{Acknowledgments}

We thank Olivier Mattelaer for useful discussions on the {\mg} implementation.
This work was supported by the Swedish Research Council (contract number
2016-05996, 
as well as the European Union's Horizon 2020 research and
innovation programme (grant agreement No 668679). 
This work has also been supported in part by the European
Union’s Horizon 2020 research and innovation programme as part of the
Marie Skłodowska-Curie Innovative Training Network MCnetITN3 (grant
agreement no. 722104).

\bibliography{MGChiralityFlow}

\providecommand{\href}[2]{#2}\begingroup\raggedright\begin{thebibliography}{10}

\bibitem{Gleisberg:2008fv}
T.~Gleisberg and S.~Hoeche, \emph{{Comix, a new matrix element generator}},
  \href{https://doi.org/10.1088/1126-6708/2008/12/039}{\emph{JHEP} {\bfseries
  12} (2008) 039} [\href{https://arxiv.org/abs/0808.3674}{{\ttfamily
  0808.3674}}].

\bibitem{Alwall:2014hca}
J.~Alwall, R.~Frederix, S.~Frixione, V.~Hirschi, F.~Maltoni, O.~Mattelaer
  et~al., \emph{{The automated computation of tree-level and next-to-leading
  order differential cross sections, and their matching to parton shower
  simulations}}, \href{https://doi.org/10.1007/JHEP07(2014)079}{\emph{JHEP}
  {\bfseries 07} (2014) 079} [\href{https://arxiv.org/abs/1405.0301}{{\ttfamily
  1405.0301}}].

\bibitem{Sjostrand:2014zea}
T.~Sj\"ostrand, S.~Ask, J.~R. Christiansen, R.~Corke, N.~Desai, P.~Ilten
  et~al., \emph{{An introduction to PYTHIA 8.2}},
  \href{https://doi.org/10.1016/j.cpc.2015.01.024}{\emph{Comput. Phys. Commun.}
  {\bfseries 191} (2015) 159}
  [\href{https://arxiv.org/abs/1410.3012}{{\ttfamily 1410.3012}}].

\bibitem{Bellm:2015jjp}
J.~Bellm et~al., \emph{{Herwig 7.0/Herwig++ 3.0 release note}},
  \href{https://doi.org/10.1140/epjc/s10052-016-4018-8}{\emph{Eur. Phys. J. C}
  {\bfseries 76} (2016) 196}
  [\href{https://arxiv.org/abs/1512.01178}{{\ttfamily 1512.01178}}].

\bibitem{Sherpa:2019gpd}
{\scshape Sherpa} collaboration, E.~Bothmann et~al., \emph{{Event Generation
  with Sherpa 2.2}},
  \href{https://doi.org/10.21468/SciPostPhys.7.3.034}{\emph{SciPost Phys.}
  {\bfseries 7} (2019) 034} [\href{https://arxiv.org/abs/1905.09127}{{\ttfamily
  1905.09127}}].

\bibitem{Murayama:1992gi}
H.~Murayama, I.~Watanabe and K.~Hagiwara, \emph{{HELAS: HELicity amplitude
  subroutines for Feynman diagram evaluations}},  tech. rep., 1, 1992.

\bibitem{DeCausmaecker:1981jtq}
P.~De~Causmaecker, R.~Gastmans, W.~Troost and T.~T. Wu, \emph{{Multiple
  Bremsstrahlung in Gauge Theories at High-Energies. 1. General Formalism for
  Quantum Electrodynamics}},
  \href{https://doi.org/10.1016/0550-3213(82)90488-6}{\emph{Nucl. Phys.}
  {\bfseries B206} (1982) 53}.

\bibitem{Berends:1981rb}
F.~A. Berends, R.~Kleiss, P.~De~Causmaecker, R.~Gastmans and T.~T. Wu,
  \emph{{Single Bremsstrahlung Processes in Gauge Theories}},
  \href{https://doi.org/10.1016/0370-2693(81)90685-7}{\emph{Phys. Lett.}
  {\bfseries 103B} (1981) 124}.

\bibitem{Berends:1981uq}
F.~A. Berends, R.~Kleiss, P.~De~Causmaecker, R.~Gastmans, W.~Troost and T.~T.
  Wu, \emph{{Multiple Bremsstrahlung in Gauge Theories at High-Energies. 2.
  Single Bremsstrahlung}},
  \href{https://doi.org/10.1016/0550-3213(82)90489-8}{\emph{Nucl. Phys.}
  {\bfseries B206} (1982) 61}.

\bibitem{DeCausmaecker:1981wzb}
P.~De~Causmaecker, R.~Gastmans, W.~Troost and T.~T. Wu, \emph{{Helicity
  Amplitudes for Massless QED}},
  \href{https://doi.org/10.1016/0370-2693(81)91025-X}{\emph{Phys. Lett.}
  {\bfseries 105B} (1981) 215}.

\bibitem{Berends:1983ez}
{\scshape CALKUL} collaboration, F.~A. Berends, R.~Kleiss, P.~de~Causmaecker,
  R.~Gastmans, W.~Troost and T.~T. Wu, \emph{{Multiple Bremsstrahlung in Gauge
  Theories at High-energies. 3. Finite Mass Effects in Collinear Photon
  Bremsstrahlung}},
  \href{https://doi.org/10.1016/0550-3213(84)90254-2}{\emph{Nucl. Phys.}
  {\bfseries B239} (1984) 382}.

\bibitem{Kleiss:1984dp}
R.~Kleiss, \emph{{The Cross-section for $e^+ e^- \to e^+ e^- e^+ e^-$}},
  \href{https://doi.org/10.1016/0550-3213(84)90197-4}{\emph{Nucl. Phys.}
  {\bfseries B241} (1984) 61}.

\bibitem{Berends:1984gf}
F.~A. Berends, P.~H. Daverveldt and R.~Kleiss, \emph{{Complete Lowest Order
  Calculations for Four Lepton Final States in electron-Positron Collisions}},
  \href{https://doi.org/10.1016/0550-3213(85)90541-3}{\emph{Nucl. Phys.}
  {\bfseries B253} (1985) 441}.

\bibitem{Gunion:1985bp}
J.~F. Gunion and Z.~Kunszt, \emph{{Four jet processes: gluon-gluon scattering
  to nonidentical quark - anti-quark pairs}},
  \href{https://doi.org/10.1016/0370-2693(85)90879-2}{\emph{Phys. Lett.}
  {\bfseries 159B} (1985) 167}.

\bibitem{Gunion:1985vca}
J.~F. Gunion and Z.~Kunszt, \emph{{Improved Analytic Techniques for Tree Graph
  Calculations and the G g q anti-q Lepton anti-Lepton Subprocess}},
  \href{https://doi.org/10.1016/0370-2693(85)90774-9}{\emph{Phys. Lett.}
  {\bfseries 161B} (1985) 333}.

\bibitem{Kleiss:1985yh}
R.~Kleiss and W.~J. Stirling, \emph{{Spinor Techniques for Calculating p anti-p
  {$\to$} W{$^{\pm}$} / Z$^0$ + Jets}},
  \href{https://doi.org/10.1016/0550-3213(85)90285-8}{\emph{Nucl. Phys.}
  {\bfseries B262} (1985) 235}.

\bibitem{Hagiwara:1985yu}
K.~Hagiwara and D.~Zeppenfeld, \emph{{Helicity Amplitudes for Heavy Lepton
  Production in e+ e- Annihilation}},
  \href{https://doi.org/10.1016/0550-3213(86)90615-2}{\emph{Nucl. Phys.}
  {\bfseries B274} (1986) 1}.

\bibitem{Kleiss:1986ct}
R.~Kleiss, \emph{{Hard Bremsstrahlung Amplitudes for $e^+ e^-$ Collisions With
  Polarized Beams at {LEP} / {SLC} Energies}},
  \href{https://doi.org/10.1007/BF01552550}{\emph{Z. Phys.} {\bfseries C33}
  (1987) 433}.

\bibitem{Kleiss:1986qc}
R.~Kleiss and W.~J. Stirling, \emph{{Cross-sections for the Production of an
  Arbitrary Number of Photons in Electron - Positron Annihilation}},
  \href{https://doi.org/10.1016/0370-2693(86)90454-5}{\emph{Phys. Lett.}
  {\bfseries B179} (1986) 159}.

\bibitem{Xu:1986xb}
Z.~Xu, D.-H. Zhang and L.~Chang, \emph{{Helicity Amplitudes for Multiple
  Bremsstrahlung in Massless Nonabelian Gauge Theories}},
  \href{https://doi.org/10.1016/0550-3213(87)90479-2}{\emph{Nucl. Phys.}
  {\bfseries B291} (1987) 392}.

\bibitem{Gastmans:1987qz}
{\scshape CALKUL} collaboration, R.~Gastmans, F.~A. Berends, D.~Danckaert,
  P.~De~Causmaecker, R.~Kleiss, W.~Troost et~al., \emph{{New techniques and
  results in gauge theory calculations}},  in \emph{{Electroweak effects at
  high-energies. Proceedings, 1st Europhysics study conference, Erice, Italy,
  February 1-12, 1983}}, pp.~599--609, 1987.

\bibitem{Schwinn:2005pi}
C.~Schwinn and S.~Weinzierl, \emph{{Scalar diagrammatic rules for Born
  amplitudes in QCD}},
  \href{https://doi.org/10.1088/1126-6708/2005/05/006}{\emph{JHEP} {\bfseries
  05} (2005) 006} [\href{https://arxiv.org/abs/hep-th/0503015}{{\ttfamily
  hep-th/0503015}}].

\bibitem{Farrar:1983wk}
G.~R. Farrar and F.~Neri, \emph{{How to Calculate 35640 O ($\alpha^5$) Feynman
  Diagrams in Less Than an Hour}},
  \href{https://doi.org/10.1016/0370-2693(85)90526-X,
  10.1016/0370-2693(83)91074-2}{\emph{Phys. Lett.} {\bfseries 130B} (1983)
  109}.

\bibitem{Berends:1987cv}
F.~A. Berends and W.~Giele, \emph{{The Six Gluon Process as an Example of
  Weyl-Van Der Waerden Spinor Calculus}},
  \href{https://doi.org/10.1016/0550-3213(87)90604-3}{\emph{Nucl. Phys.}
  {\bfseries B294} (1987) 700}.

\bibitem{Berends:1987me}
F.~A. Berends and W.~T. Giele, \emph{{Recursive Calculations for Processes with
  n Gluons}}, \href{https://doi.org/10.1016/0550-3213(88)90442-7}{\emph{Nucl.
  Phys.} {\bfseries B306} (1988) 759}.

\bibitem{Berends:1988yn}
F.~A. Berends, W.~T. Giele and H.~Kuijf, \emph{{Exact Expressions for Processes
  Involving a Vector Boson and Up to Five Partons}},
  \href{https://doi.org/10.1016/0550-3213(89)90242-3}{\emph{Nucl. Phys.}
  {\bfseries B321} (1989) 39}.

\bibitem{Berends:1988zn}
F.~A. Berends and W.~T. Giele, \emph{{Multiple Soft Gluon Radiation in Parton
  Processes}}, \href{https://doi.org/10.1016/0550-3213(89)90398-2}{\emph{Nucl.
  Phys.} {\bfseries B313} (1989) 595}.

\bibitem{Berends:1989hf}
F.~A. Berends, W.~T. Giele and H.~Kuijf, \emph{{Exact and Approximate
  Expressions for Multi - Gluon Scattering}},
  \href{https://doi.org/10.1016/0550-3213(90)90225-3}{\emph{Nucl. Phys.}
  {\bfseries B333} (1990) 120}.

\bibitem{Dittmaier:1993bj}
S.~Dittmaier, \emph{{Full O(alpha) radiative corrections to high-energy Compton
  scattering}}, \href{https://doi.org/10.1016/0550-3213(94)90139-2}{\emph{Nucl.
  Phys.} {\bfseries B423} (1994) 384}
  [\href{https://arxiv.org/abs/hep-ph/9311363}{{\ttfamily hep-ph/9311363}}].

\bibitem{Dittmaier:1998nn}
S.~Dittmaier, \emph{{Weyl-van der Waerden formalism for helicity amplitudes of
  massive particles}},
  \href{https://doi.org/10.1103/PhysRevD.59.016007}{\emph{Phys. Rev.}
  {\bfseries D59} (1998) 016007}
  [\href{https://arxiv.org/abs/hep-ph/9805445}{{\ttfamily hep-ph/9805445}}].

\bibitem{Weinzierl:2005dd}
S.~Weinzierl, \emph{{Automated computation of spin- and colour-correlated Born
  matrix elements}},
  \href{https://doi.org/10.1140/epjc/s2005-02467-6}{\emph{Eur. Phys. J.}
  {\bfseries C45} (2006) 745}
  [\href{https://arxiv.org/abs/hep-ph/0510157}{{\ttfamily hep-ph/0510157}}].

\bibitem{Mangano:1990by}
M.~L. Mangano and S.~J. Parke, \emph{{Multiparton amplitudes in gauge
  theories}}, \href{https://doi.org/10.1016/0370-1573(91)90091-Y}{\emph{Phys.
  Rept.} {\bfseries 200} (1991) 301}
  [\href{https://arxiv.org/abs/hep-th/0509223}{{\ttfamily hep-th/0509223}}].

\bibitem{Dixon:1996wi}
L.~J. Dixon, \emph{{Calculating scattering amplitudes efficiently}},  in
  \emph{{QCD and beyond. Proceedings, Theoretical Advanced Study Institute in
  Elementary Particle Physics, TASI-95, Boulder, USA, June 4-30, 1995}},
  pp.~539--584, 1996, \href{https://arxiv.org/abs/hep-ph/9601359}{{\ttfamily
  hep-ph/9601359}}.

\bibitem{Dreiner:2008tw}
H.~K. Dreiner, H.~E. Haber and S.~P. Martin, \emph{{Two-component spinor
  techniques and Feynman rules for quantum field theory and supersymmetry}},
  \href{https://doi.org/10.1016/j.physrep.2010.05.002}{\emph{Phys. Rept.}
  {\bfseries 494} (2010) 1} [\href{https://arxiv.org/abs/0812.1594}{{\ttfamily
  0812.1594}}].

\bibitem{Elvang:2013cua}
H.~Elvang and Y.-t. Huang, \emph{{Scattering Amplitudes}},
  \href{https://arxiv.org/abs/1308.1697}{{\ttfamily 1308.1697}}.

\bibitem{Dixon:2013uaa}
L.~J. Dixon, \emph{{A brief introduction to modern amplitude methods}},  in
  \emph{{Proceedings, 2012 European School of High-Energy Physics (ESHEP 2012):
  La Pommeraye, Anjou, France, June 06-19, 2012}}, pp.~31--67, 2014,
  \href{https://arxiv.org/abs/1310.5353}{{\ttfamily 1310.5353}},
  \href{https://doi.org/10.5170/CERN-2014-008.31}{DOI}.

\bibitem{Alnefjord:2020xqr}
J.~Alnefjord, A.~Lifson, C.~Reuschle and M.~Sjodahl, \emph{{The chirality-flow
  formalism for the standard model}},
  \href{https://doi.org/10.1140/epjc/s10052-021-09055-2}{\emph{Eur. Phys. J. C}
  {\bfseries 81} (2021) 371}
  [\href{https://arxiv.org/abs/2011.10075}{{\ttfamily 2011.10075}}].

\bibitem{Lifson:2020oll}
A.~Lifson, C.~Reuschle and M.~Sj\"odahl, \emph{{Introducing the Chirality-flow
  Formalism}}, \href{https://doi.org/10.5506/APhysPolB.51.1547}{\emph{Acta
  Phys. Polon. B} {\bfseries 51} (2020) 1547}.

\bibitem{Lifson:2020pai}
A.~Lifson, C.~Reuschle and M.~Sjodahl, \emph{{The chirality-flow formalism}},
  \href{https://doi.org/10.1140/epjc/s10052-020-8260-8}{\emph{Eur. Phys. J. C}
  {\bfseries 80} (2020) 1006}
  [\href{https://arxiv.org/abs/2003.05877}{{\ttfamily 2003.05877}}].

\bibitem{Alnefjord:2021yks}
J.~Alnefjord, A.~Lifson, C.~Reuschle and M.~Sjodahl, \emph{{A Brief Look at the
  Chirality-Flow Formalism for Standard Model Amplitudes}},
  \href{https://doi.org/10.22323/1.397.0160}{\emph{PoS} {\bfseries LHCP2021}
  (2021) 160} [\href{https://arxiv.org/abs/2110.04125}{{\ttfamily
  2110.04125}}].

\bibitem{Degrande:2011ua}
C.~Degrande, C.~Duhr, B.~Fuks, D.~Grellscheid, O.~Mattelaer and T.~Reiter,
  \emph{{UFO - The Universal FeynRules Output}},
  \href{https://doi.org/10.1016/j.cpc.2012.01.022}{\emph{Comput. Phys. Commun.}
  {\bfseries 183} (2012) 1201}
  [\href{https://arxiv.org/abs/1108.2040}{{\ttfamily 1108.2040}}].

\bibitem{deAquino:2011ub}
P.~de~Aquino, W.~Link, F.~Maltoni, O.~Mattelaer and T.~Stelzer, \emph{{ALOHA:
  Automatic libraries of helicity amplitudes for Feynman diagram
  computations}},
  \href{https://doi.org/10.1016/j.cpc.2012.05.004}{\emph{Computer Physics
  Communications} {\bfseries 183} (2012) 2254–2263}.

\bibitem{Mattelaer:2021xdr}
O.~Mattelaer and K.~Ostrolenk, \emph{{Speeding up MadGraph5\_aMC@NLO}},
  \href{https://doi.org/10.1140/epjc/s10052-021-09204-7}{\emph{Eur. Phys. J. C}
  {\bfseries 81} (2021) 435}
  [\href{https://arxiv.org/abs/2102.00773}{{\ttfamily 2102.00773}}].

\bibitem{Kleiss:1985gy}
R.~Kleiss, W.~J. Stirling and S.~D. Ellis, \emph{{A New Monte Carlo Treatment
  of Multiparticle Phase Space at High-energies}},
  \href{https://doi.org/10.1016/0010-4655(86)90119-0}{\emph{Comput. Phys.
  Commun.} {\bfseries 40} (1986) 359}.

\bibitem{Nethercote:2007vaf}
N.~Nethercote and J.~Seward, \emph{Valgrind: a framework for heavyweight
  dynamic binary instrumentation},  in \emph{PLDI '07}, 2007.

\bibitem{Weidendorfer:2004ccs}
J.~Weidendorfer, M.~Kowarschik and C.~Trinitis, \emph{A tool suite for
  simulation based analysis of memory access behavior},  vol.~3038,
  pp.~440--447, 06, 2004,
  \href{https://doi.org/10.1007/978-3-540-24688-6_58}{DOI}.

\end{thebibliography}\endgroup

\end{document}